\documentstyle[11pt,epsfig,psfig]{article}
\columnwidth\textwidth
\newcommand{\avrho}{\mbox{$\langle \rho\rangle$}}
\newcommand{\avrhot}{\mbox{$\langle \rho(t)\rangle$}}
\newcommand{\avrhop}{\mbox{$\langle\rho'\rangle$}}
\newcommand{\avrhopt}{\mbox{$\langle\rho'(t)\rangle$}}

\def\gsim{\;
\raise0.3ex\hbox{$>$\kern-0.75em\raise-1.1ex\hbox{$\sim$}}\;
}
\def\lsim{\;
\raise0.3ex\hbox{$<$\kern-0.75em\raise-1.1ex\hbox{$\sim$}}\;
\ref}

\begin{document}

\begin{flushright}
IFIC/FTUV-9923 \\
\end{flushright}

\vspace{0.5cm}
\begin{center}

{\bf \Large Influence of solar chaotic  magnetic fields on neutrino oscillations}\\
\end{center}

\begin{center}
E. Torrente-Lujan\footnote{Presenting work made on collaboration with
V.B. Semikoz.
To be published in the proceedings of 2nd ICRA Network Workshop: The Chaotic Universe: Theory and observations, Rome, Italy, 1-5 Feb 1999.
}
\\
Dept. de F\'{\i}sica Te\`orica. Univ. de Val\`encia,
46100 Burjassot, Val\`encia, Spain\\
{E-mail address: e.torrente@cern.ch}

\end{center}

\begin{abstract}
 We consider the effect of a random magnetic field in the
 convective zone of the Sun 
 on resonant neutrino spin-flavour oscillations.
The expected signals in the different experiments
 (SK,GALLEX-SAGE,Homestake) are obtained as a function of
 the level of noise, regular magnetic field and neutrino mixing
 parameters. Previous results obtained for small mixing and
 ad-hoc regular magnetic profiles are reobtained. We find that
 MSW regions are stable up to very large levels of noise
 (P=0.7-0.8) and they are acceptable from the point of view of
 antineutrino production.
For strong noise any parameter region 
$(\Delta m^2, \sin^2 2\theta)$ is excluded: 
this model of noisy magnetic field 
 is not compatible with particle physics solutions to the SNP. 
\end{abstract}

\vspace{0.5cm}

\section{Introduction}

A neutrino transition magnetic moment can account, 
both for the observed 
deficiency of the solar neutrino flux and the time variations of the signal. 
The overall deficit is caused by the suppression 
of the neutrino energy 
spectrum.
The time dependence may be caused by time variations of the convective 
magnetic 
field  of the Sun \cite{akh4}.
In transverse magnetic fields, neutrinos with transition magnetic moments will 
experience spin and flavor rotation simultaneously (resonant 
spin-flavor precession, 
RSFP). 
The observation of electron antineutrinos from the Sun would lead to the conclusion that the 
neutrinos are Majorana particles. There are however stringent bounds on the presence of solar electron antineutrinos 
coming from solar $^8$B neutrinos \cite{fio1,fio2}.

Magnetic fields measured on the surface of the Sun are
 weaker than those ones within the interior of the convective zone. 
The mean field value over the solar disc is  of the order of 1 G and in the 
solar spots magnetic field strength reaches $\ 1$ kG. In magnetic 
hydrodynamics, MHD, one can explain
such fields in a self-consistent way if these fields are generated 
by the existence of a  dynamo effect at the 
bottom of the convective zone.
In this region  the strength of small scale 
 regular magnetic fields could  reach a value of 100 kG.
These fields propagate through the  convective zone and photosphere 
decreasing in  strength 
while increasing in the scale giving traces in from of loops in bipolar
 active regions (solar spots).

Large-scale toroidal magnetic field created by a dynamo mechanism in
 convective zone 
typically have strengths even less than small-scale r.m.s fields near the bottom of convective zone. This is 
the main reason why one should  consider 
 neutrino propagation in the random magnetic field of the Sun. 
The ratio of the r.m.s. random field and the regular (toroidal) 
field may be about $\sim$ 20-50, therefore the 
problem of RSFP neutrino propagation  in noisy magnetic field seems to be important. 
Estimations for the ratio of rms fields to regular field  are necessarily very 
rough.
In textbooks \cite{vai1,vai2,park1,park2}  
we find the conservative ratio 
$\eta\equiv\langle\tilde{B}^2\rangle/B^2_0\sim 1.$ 
In more elaborate models  the ratio of magnetic energy densities is given by the magnetic 
Reynolds number, 
$\langle\tilde{B}^2\rangle/B^2_0\sim R_m^{\sigma},$ 
which may be  much bigger than unity for plasma
 with large conductivity. Here 
$\sigma>0$ is a topology index \cite{vai1,vai2,Cataneo,Diamond}.

The effect of random magnetic fields in RSFP solutions to the solar neutrino problem, SNP, and 
antineutrino production has been explored  previously for simplified
 models \cite{tor3,tor4}.
In this work we will deal with the complete problem, 
we will present calculations of neutrino spin flavor conversions in 
presence of matter and magnetic field. 
The magnetic field will have two well differentiated components. 
The first component  will be   a theoretically 
motivated solar magnetic field profile:  solution to the  static 
magnetic hydrodynamic equations.
As a second component, we include 
  a layer of magnetic noise generated at the bottom of the convective 
zone, we have justified that the level of noise
in this region can  certainly  be very high.

\section{The Master equation}
We consider conversions $\nu_{eL,R}\to \nu_{aL,R}$, $a = \mu$ or
$\tau$, (for definiteness we will refer to  $\mu$ for the rest of this work) for two neutrino flavors obeying the 
following master evolution equation
\begin{eqnarray}
i \partial_t 
\pmatrix{\nu_{eL}\cr\overline{\nu}_{eR}\cr\nu_{\mu L}\cr\overline{\nu}_{\mu R}} 
&= &
\pmatrix{V_e -c_2\delta &0&s_2\delta &\mu B_{\perp}^+(t)\cr 
0& - V_e - c_2\delta & - \mu B_{\perp}^-(t)& s_2\delta \cr
s_2\delta & - \mu B_{\perp}^+(t)&V_{\mu} + c_2\delta &0\cr
\mu B_{\perp}^-(t)&s_2\delta &0& - V_{\mu} + c_2\delta}
\pmatrix{\nu_{eL}\cr\overline{\nu}_{eR}\cr\nu_{\mu L}\cr\overline{\nu}_{\mu R}}, 
\label{e1001}
\label{master}
\end{eqnarray} 
where $c_2 = \cos 2\theta$, $s_2 = \sin (2\theta)$ and $\delta = \Delta m^2/4E$ are the neutrino 
mixing parameters; $\mu \equiv \mu_{12}$ is the neutrino active-active
transition magnetic moment, 
\begin{eqnarray}
B_{\perp}^\pm(t)& =& B_{0\perp}^\pm(t) + \tilde{B}_{\perp}^\pm(t)
\label{e7701}
\end{eqnarray}
 is the magnetic field component which is perpendicular 
to the neutrino trajectory in the Sun. The quantities  
$V_{e,\mu}(t)$  
are the neutrino 
vector potentials for $\nu_{eL}$ and $\nu_{\mu L}$ in the Sun given 
by the abundances of the electron  and 
neutron  components and 
by the SSM density profile \cite{BP95}.

The transverse magnetic field  $B_{\perp}(t)$ 
appearing in Eq. (\ref{e7701}) is 
given by
$ B_{\perp}^{\pm}
(t) \equiv B_x(t) \pm i B_y(t)\equiv |B_{\perp}(t)| Exp (\pm i\Phi (t)).$
The nature and magnitude for the 
regular  ($B_{0}$) and chaotic parts ($\tilde{B}$) of the 
magnetic field will be the subject of the next section. 
In our calculations we will consider always  the product $\mu B$. 
Expected values of $B\approx 1-100$ kG in the Sun convective zone and 
$\mu=10^{-11} \mu_B$ would give an expected range for the 
 product $\mu B\approx 10^{-8}-10^{-6} \ \mu_B G\approx 5.6\ 10^{-17}-10^{-15} $ eV  or in the 
practical units which will be used 
 throughout
 this work $\mu B\approx \ 0.1-10.0\ \mu_{11} B_4$.

\section{Solar magnetic fields}

\subsection{Random magnetic fields}
The r.m.s. random component $\sqrt{\langle \tilde{B}^2(t)\rangle}$ 
can be comparable in magnitude with the regular one, $B_0(t)$, and maybe even
much stronger than $B_0$, 
if  a large magnetic Reynolds number $R_m$ leads to the effective dynamo 
enhancement of small-scale (random) magnetic fields.

Let us give some 
simple estimates of the magnetic Reynolds number $R_m = lv/\nu_m$
in the convective zone for fully ionized hydrogen plasma ($T\gg I_H\sim 13.6~eV\sim 10^5~K$). 
Here $l\sim 10^8~cm$ is the size of eddy (of the order of magnitude of a granule size) with the 
turbulent velocity inside of it
$v\sim v_A\sim 10^5~cm/s$ where $v_A = B_0/\sqrt{4\pi \rho}$ is the Alfven velocity for MHD 
plasma,
$B_0$ is a large-scale field in the zone and $\rho$ is the matter density (in $g/cm^3$.

The magnetic diffusion coefficient, $\nu_m = c^2/4\pi \sigma_{cond}$,  
 appears in  the diffusion term  of 
the Faraday equation:
\begin{equation}
\frac{\partial \vec{B}(t)}{\partial t} = rot [\vec{v}\times \vec{B}(t)] + \nu_m \Delta \vec{B}~.
\label{Faradey}
\end{equation} 
Here $c$ is the light speed. 
The conductivity of the hydrogen plasma is $\sigma_{cond} = \omega_{pl}^2/4\pi 
\nu_{ep}$, where 
$\omega_{pl} = \sqrt{4\pi e^2n_e/m_e} = 5.65\times 10^4\sqrt{n_e}~s^{-1}$ 
is the plasma (Langmuir) frequency;
$\nu_{ep} = 50n_e/T^{3/2}~s^{-1}$ is the electron-proton collision frequency, 
the electron 
density $n_e( = n_p)$ ($cm^{-3}$)
and the temperature $T$ (K).

Thus we find that the magnetic diffusion coefficient 
$\nu_m\simeq 10^{13}(T/1~K)^{-3/2}\ {\rm cm}^2{\rm s}^{-1}$
 does not depend 
on the charge density $n_e$ and it is very small  relative to the 
product $v l$  for hot plasma 
$T\geq 10^5~K\gg 1~K$. 
From the 
comparison of the first and second terms in the r.h.s. of the 
Faraday equation Eq. (\ref{Faradey}) we find that $v/l\gg \nu_m/l^2$,
 or $\nu_m\ll vl\sim 10^{13}~cm^2s^{-1}$ since $T/1~K\gg 1$.
This means that the magnetic field in the Sun is mainly {\it frozen-in}. 
Neglecting 
the second term in Eq. (\ref{Faradey}) and using the Maxwell equation 
$rot \vec{E}= - c^{-1}(\partial \vec{B}/\partial t)$
we obtain the condition for frozen-in field:
the Lorentz force vanishes, 
$\sim (\vec {E} + [\vec{v}\times \vec{B}]/c)\simeq 0$ 
but the current 
$\vec{j} = \sigma_{cond}(\vec {E} + [\vec{v}\times \vec{B}]/c)$
remains finite if the conductivity is large, $\sigma_{cond}\to \infty$.

The magnetic Reynolds number 
$$R_m = lv\omega_{pl}^2/(c^2\nu_{ep})\simeq lv\times 10^{-13}(T/1~K)^{3/2}~cm^2s^{-1}$$ 
is huge if we substitute the estimate
$lv\sim 10^{13}~cm^2s^{-1}$ given above. A large value for the 
Reynolds number  is a necessary condition for the existence of an effective  
dynamo enhancement in the convective zone.

The 
random magnetic field component in the Sun, $\tilde{B}(t)$, 
will be  described in general by 
an arbitrary, phenomenological, correlator of the form
$\langle \tilde{B}(t)\tilde{B}(t^{'})\rangle = 
\langle \tilde {B}^2\rangle f(t - t^{'}).$
 We will assume that 
the strength of
the r.m.s. field squared $\langle \tilde{B}^2\rangle = \eta B_0^2$ 
is
parametrized by the 
dimensionless parameter $\eta = R_m^{\sigma}>1$ , which it  can be, in general, much bigger than unity, 
$\eta\gg 1$. 
Actually, the estimation of the quantity $\eta$, 
the ratio of rms fields to regular field,
for the solar 
convective zone (and other cosmic dynamos) is the matter 
of current scientific discussions.
The most conservative estimate, simply based on
 equipartition, is $\eta=constant$. 
According to  direct observations of galactic
magnetic field presumably driven by a dynamo, 
$\eta \simeq 1.8 $ \cite{Ruzmaikin}. 
A more developed theory of equipartition
gives  $\eta \simeq 4 \pi  \ln R_m$ 
(see Ref. \cite{Zeldovich}). 
Let us note that this estimate is considered now as very conservative: 
more detailed theories of MHD turbulence yield 
estimates like $\eta \sim \sqrt{R_m} $
\cite{Cataneo,Diamond}.

The correlator function $f(t)$ is unknown a priori but it
 takes the particular
$\delta$-correlator form
$f(t) = L_0\delta (t)$ 
if the correlation length (for two neighboring magnetic field domains) is much less than the 
neutrino oscillation length, $L_0\ll l_{osc}$.   
$L_0$ can be considered  a free parameter
 ranging in the interval $10^0-10^4$ km.
In the averaged evolution equations it only appears  the 
product  $\eta L_0$. Thus  
in what follows we will present our results as a function of the
quantity P which  is a simple function of such a product:
\begin{eqnarray}
P&\equiv &\frac{1}{2}\left (1+ \exp(-\gamma)\right ),\quad
\gamma\equiv\frac{4}{3} \Omega^2 \Delta t\equiv 
\frac{4}{3} \eta L_0 (\mu B_{0})^2 \Delta t.
\label{e2003b}
\end{eqnarray}

The reason for using the quantity 
$P$ is that it  is a good approximation for the 
 depolarization that the presence of noise induces in the averaged 
neutrino density matrix. 
 $\Delta t$ is the distance over which the noise is acting.
We have  supposed in our computations that the noise is only effective  in a thin layer with thickness 
$\Delta t=0.1\ R_\odot$ starting  at $r=0.7\ R_\odot$, the
bottom of the convective zone. 
 In Table (1) of Ref.(\cite{torrente}) the quantities 
 $\sqrt{\langle B_0^2\rangle}$ and  $\eta$ are  computed for a given $P$ 
supposing the  reasonable value  $L_0=1000$ Km.
For example for 
$P=0.95$,  $\sqrt{ \langle B_0^2\rangle}=10$ and  
$\eta(\mu B=1)=100$ (all $\mu B$ are given in $\mu_{11} B_4$ units).

\subsection{Regular large-scale magnetic field in the Sun. The twisting field}
In this work, we will apply for the neutrino conversions described by our master 
equation Eq. (\ref{master}) 
the self-consistent model of large-scale regular field given in Ref. \cite{kut1}. 
In this model,
the global solar magnetic field is the axisymmetric equilibrium solution of 
the MHD static equations (quiet Sun) in the spherically symmetric 
gravitational field of the Sun. The reasonable 
boundary condition  $B_0 = 0$ on the photosphere ($r=R_{\odot}$) is imposed in addition.
Any field solution to these equations and boundary conditions is a twisting field 
{\it with an, arbitrary, small or large 
number of revolutions along radius} ($k=1,2,\cdots$, 
the twist rate can be taken as a label to distinguish particular solution within the family).   
Full expressions for the 
 spherical components of the magnetic field 
can be found in Ref.(\cite{torrente}).
There is only a free parameter in this model.
The constant $K$ 
is related with
central field $B_{core}$ by:
$K = B_{core}/2(1 - \alpha R_{\odot}/\sin \alpha R_{\odot})$.
The modulus of the perpendicular component is of the form:
$B_{0\perp} = B_{core}\sin\theta  f(r)/r$
where $f(r)$ is some known function of  gentle behavior.
According to this model, the expected magnetic field at the core is 
typically only 2-3  times (or less) 
the magnetic field at the convective zone. For the 
values  that we will consider later, $B_{0.7}\approx <100-200$ kG, the values 
corresponding at the core are well below astrophysical bounds
derived from traces of these fields at solar surface.

\section{The averaged master equation}
The master Equation (\ref{e1001}) 
can be written in terms of the density matrix
$\rho(t)$ as:
\begin{eqnarray}
i\partial_t \rho&=&[H_{reg},\rho]+\mu \tilde{B}_x(t)[V_x,\rho]+\mu \tilde{B}_y(t)[V_y,\rho].
\label{e7702}
\end{eqnarray}
The elements
 of the matrices $H_0, V_x,V_y$ can be read off the Eq. (\ref{e1001}). 
The $\tilde{B}_x,\tilde{B}_y$ are the Cartesian transversal components
of the chaotic magnetic field. Vacuum mixing terms and matter terms corresponding to the SSM density profile given before and the regular  
magnetic part Hamiltonian are all included in $H_{reg}$. 
In particular, the matrices $V_{x,y}$ are Gamma-like matrices 
 in terms of the 
Pauli matrices $\sigma_{1,2}$ \cite{torrente}.

It is our objective in this section to write the differential evolution equation for the average density matrix $\langle\rho\rangle$.
We assume that the components 
 $\tilde{B}_x,\tilde{B}_y$ are statistically independent, each of them 
characterized by a delta-correlation function
($\langle \tilde{B}_{x}(t)\tilde{B}_{y}(t')\rangle = 0$) .
\begin{eqnarray}
\langle \tilde{B}_{x,y}(t)\tilde{B}_{x,y}(t')\rangle &=& \langle \tilde {B}_{x,y}^2\rangle  L_0\delta(t - t^{'}).
\end{eqnarray}
In addition, we will make an equipartition  assumption for 
each of the three cartesian components $\tilde{B}_{x,y,z}$.
The averaged evolution equation is a simple generalization 
(see Ref. \cite{tor2} for a complete derivation) 
of the well known Redfield equation
\cite{balantekin,tor20,bur1}  for two independent sources of noise 
and reads
($\Omega^2\equiv  L_0\mu^2 \langle\tilde{B}^2\rangle/2\equiv \eta L_0 
(\mu B_{0})^2/3$):
\begin{eqnarray}
i\partial_t \avrho=[H_{reg},\avrho]-i \Omega^2[V_x,[V_x,\avrho]]-i\Omega^2 
[V_y,[V_y,\avrho]].
\label{e8690}
\end{eqnarray}

It is possible to write the Eq. (\ref{e8690}) in a more evolved form. Taking into account 
 the particular form of the matrices 
$V_{x,y}$ and 
performing a rescaling of the
density matrix given by: 
$\avrhot=\exp(-4 \Omega^2  t)\avrhopt$ we finally obtain
 the desired averaged evolution equation:
\begin{eqnarray}
i\partial_t \avrhop=[H_{reg},\avrhop]+i 2\Omega^2 \left (V_x\avrhop V_x+V_y\avrhop V_y\right ).
\label{e8692}
\end{eqnarray}

It is useful however to consider the 
solution to  Eq. (\ref{e8692})  when $H_{reg}\equiv0$. 
This is the appropriate limit when dealing with extremely low $\Delta m^2$ or 
very large energies, 
for an extreme level of noise or when the 
distance over which the noise is acting is small enough to consider the evolution given by 
$H_{reg}$ negligible.
 In any other scenario it can give at least an idea 
of the general
behavior of the solutions to the full  Eq. (\ref{e8692}). 
When $H_0=0$   only the two last terms   in the equation  remain 
and an
exact simple expression is obtainable by ordinary algebraic methods.
The full 4x4 Hamiltonian decouples in 2x2 blocks.
If $P_{f,i}$ are the final  and initial probabilities 
(at the exit and at the entrance of  the noise  region) 
their   averaged counterparts  fulfill 
linear relations among them, schematically:
$Q_f^{A,B}=M Q_i^{A,B}$
with $Q^{A,B}$ any of the two dimensional vectors
$Q^A=(\langle P(\nu_{eL}\to  \nu_{eL}  )\rangle, 
\langle P(\nu_{eL}\to \tilde{\nu}_{\mu R} )\rangle )
$,
$Q^B=(\langle P(\nu_{eL}\to  \tilde{\nu}_{eR}  )\rangle, 
\langle P(\nu_{eL}\to \nu_{\mu L} )\rangle )$
and the Markovian matrix $M$:
\begin{eqnarray}
M&=&\pmatrix{P & 1-P \cr 1-P & P}
\end{eqnarray}
with $P$ defined in Eq. (\ref{e2003b}).
It can be shown that in this simple case $P$ is exactly the final polarization of the 
density matrix
(one of the eigenvalues of the matrix $M$ is equal to $2P-1$). 
In the general case with a finite $H_{reg}$ it can be shown numerically 
that the quantity $P$ still gives a reasonable approximation ($< 10\%$) to the real polarization, at least for the 
cases of interest in this work. 

\section{Results and Discussion }
The present status of the Solar neutrino problem
in Tables (1) of Refs. \cite{torrente,bah2}.
For this data,
we have calculated the expected neutrino signals in the Homestake, Ga-Ge and (Super)-Kamiokande experiments. For this objective, the time averaged 
survival and transition
probabilities have been obtained 
by numerical integration of  the ensemble averaged master equation (\ref{e7702}) for a certain regular magnetic profile $B_{0\perp}$ (see Ref.\cite{torrente}
for details).
The   free parameters of our model are four: 
$\delta=\Delta m^2/2E$,
 $s_2^2\equiv \sin^2 2\theta$,  
a noise strength parameter ($P$) and 
the product of magnetic field and moment ($\mu B_{0\perp}$ 
at 
$r=0.7 R_\odot$, 
$\theta_s=7^o$).

\begin{figure}[h]
\centering
\begin{tabular}{cc}
\epsfig{file=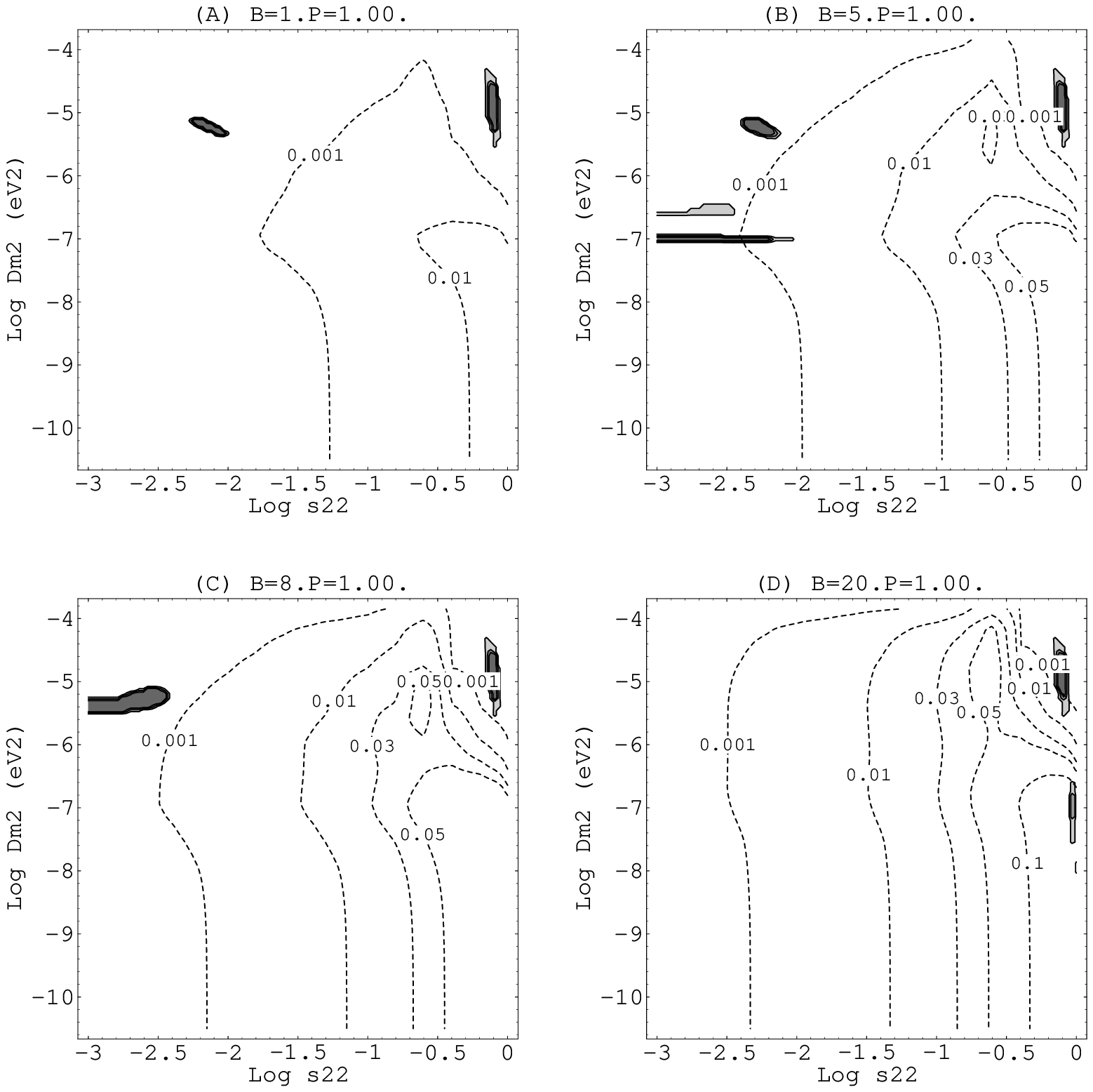,height=8cm}&
\epsfig{file=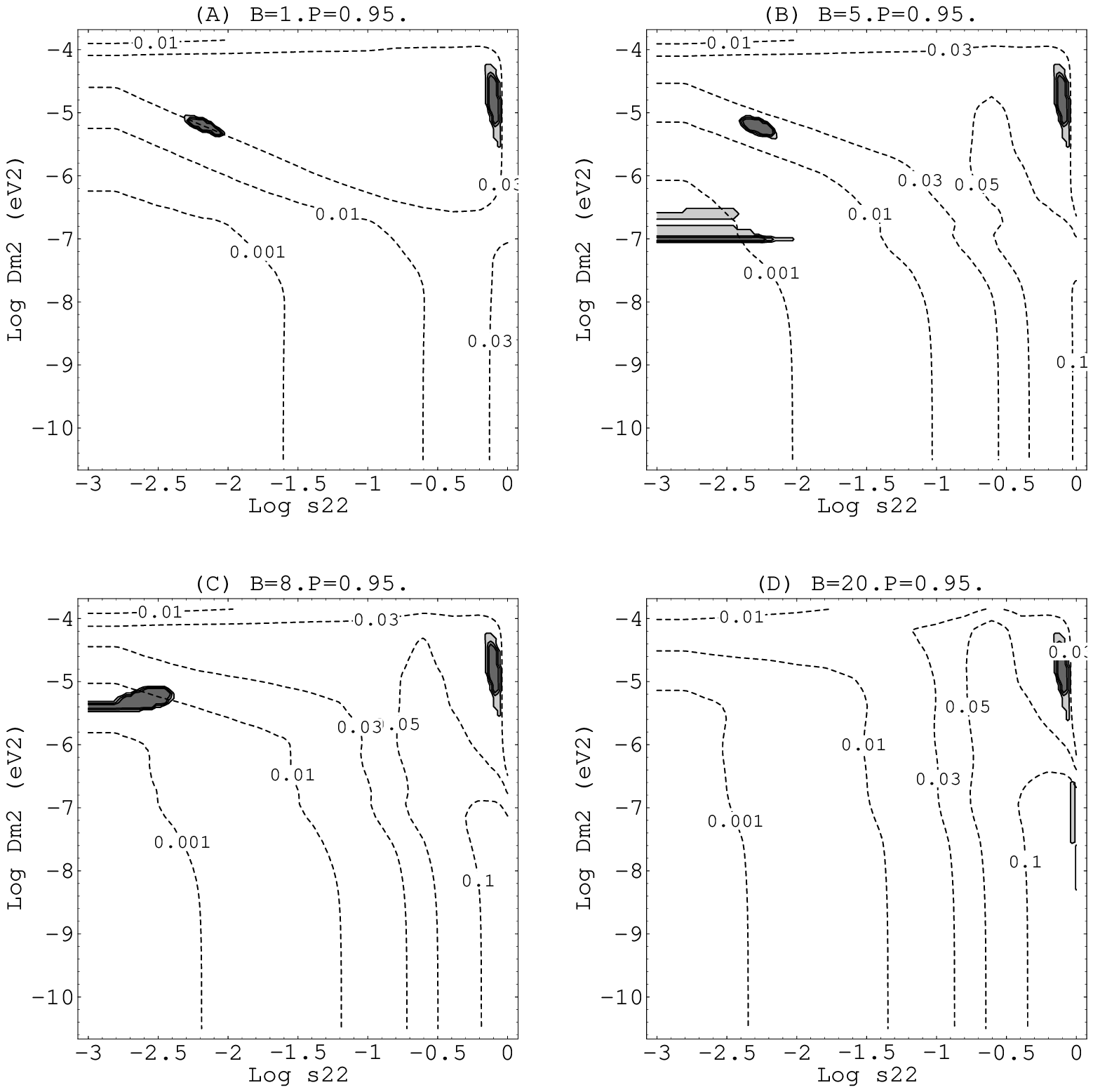,height=8cm} 
\vspace{-1.0cm}
\end{tabular}
\caption{The parameter regions consistent with the Homestake, 
Kamiokande and combined Gallium experiments in 
absence of noise (Left Block of four figures): $P=1.00$. 
(C.L.$=90\%,95\%,99\% $ (from darkest to lighter shaded areas)).
Figures  (a,b,c,d) correspond  to 
values of the regular magnetic field at the 
bottom of the convective zone: $\mu B=1,5,8,20\ \mu_{11} B_{4}$.
The electron antineutrino averaged probability, 
is represented by the dashed lines. Present 
Kamiokande bounds are 
$P_{\nu_e\overline{\nu_e}}<0.05$.
Right Block: the same for  $P=0.95$.}
\label{f14N99}
\label{f14N100}
\end{figure}

\begin{figure}[h]
\centering\hspace{0.8cm}
\begin{tabular}{cc}
\epsfig{file=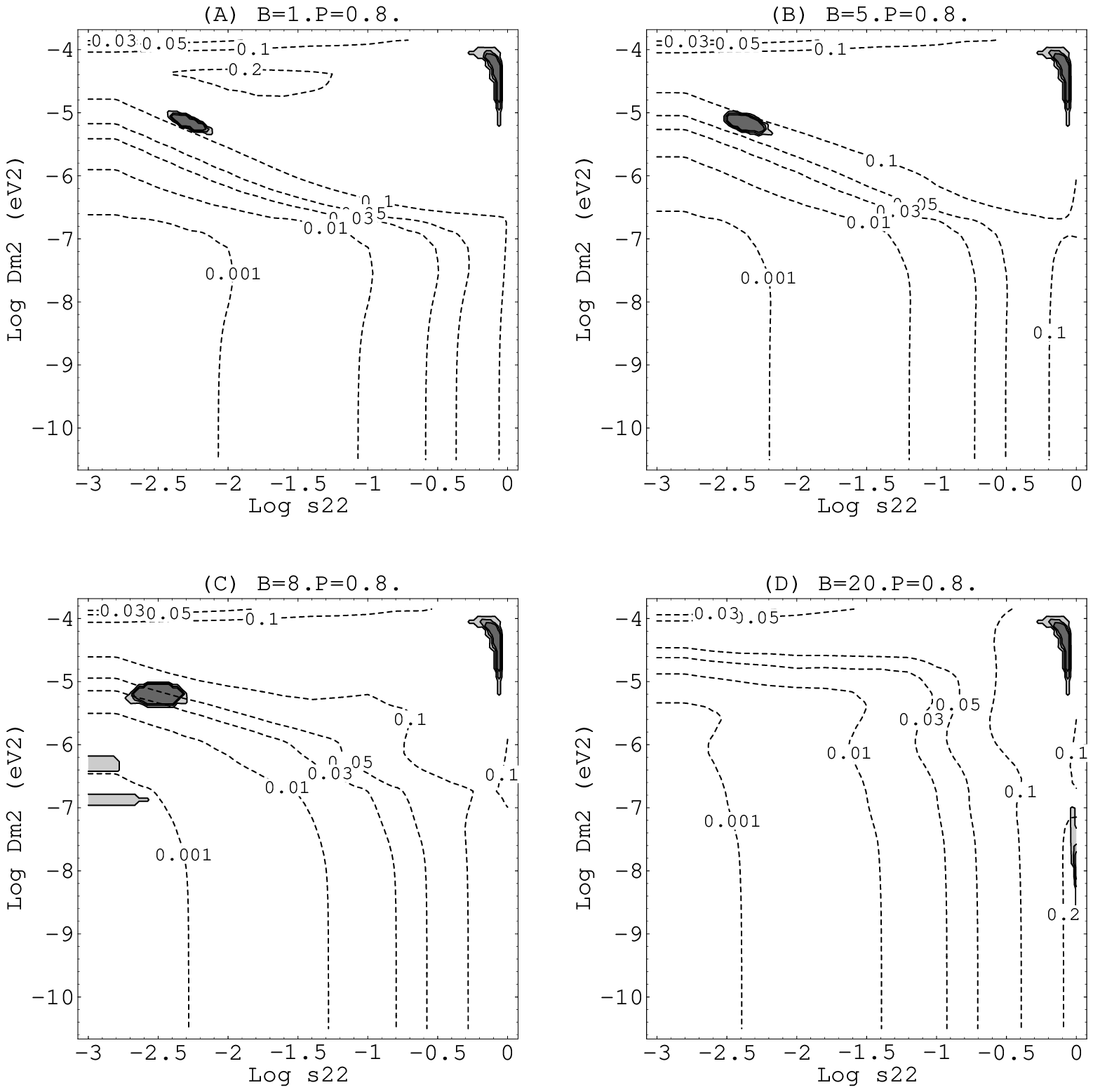,height=8cm}&
\epsfig{file=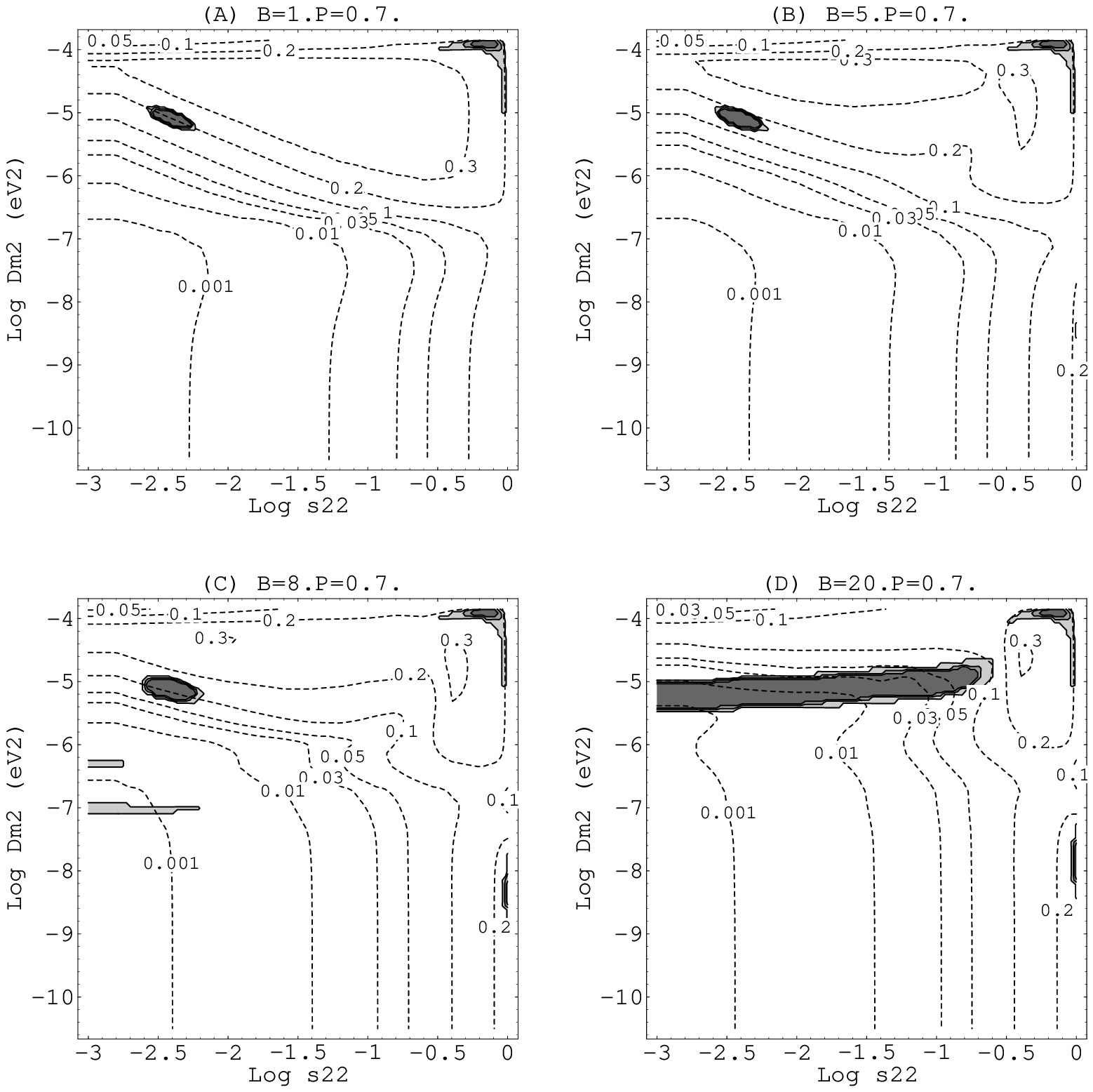,height=8cm}
\end{tabular}
\vspace{-1.0cm}
\caption{Left Block of four  Figures: As Fig. \protect\ref{f14N99} for $P=0.80$.
Right Block:  $P=0.70$.}
\label{f14N95}
\label{f14N90}
\end{figure}

In Figs. \ref{f14N100}-\ref{f14N90} 
we present the 
$(\Delta m^2, \sin^2 2\theta)$ exclusion plots 
from a combined $\chi^2$ analysis of the three experiments.
First we comment the results in complete absence of noise (P=1) which are 
represented
in Fig. \ref{f14N100} (Left block of four plots). For negligible or low regular magnetic field we observe
 the  high squared mass difference solutions 
proportioned by the matter MSW effect. As the magnitude of the magnetic 
field increases new solutions appear and disappear in a complicated manner.
The low angle solution however disappears at  high 
magnetic fields, after experiencing some distortion coming from its 
merging with newborn magnetic solutions (compare low angle allowed 
regions in Figures  (b) and (c)).
The antineutrino production (dashed lines) is in general  low  and 
Kamiokande bounds are not specially restrictive except at very 
high magnetic fields. Note in Figure (d) the 
very different behavior of the two existing allowed regions:
while the in MSW region the antineutrino production is in the $0.1-1\%$,
 compatible comfortably with Kamiokande bounds, 
the RSFP solution reach a value well above $10\%$ and is excluded by them.  
It seems apparent that there are acceptable particle solutions to the
SNP even for very large regular magnetic fields.

The pattern of the electron antineutrino probability is very different
when  a small level of noise (P=0.95, Fig. \ref{f14N99}, right block of four plots) is switched on.
For $\Delta m^2> 10^{-6}$ eV$^2$: 
the antineutrino iso-probability lines follow 
 the characteristic MSW triangular patterns in this region.
The structure of the allowed regions remain unmodified. The electron 
antineutrino yield in these regions is below the
 $5\%$ level similarly as before.
For   stronger levels of noise 
(P=0.8, Figs. \ref{f14N95})  the same comments can be said. The structure and 
position of 
the allowed regions from combined total rates are practically
unmodified but the antineutrino yield impose strong restrictions. For 
an antineutrino probability smaller than $3\%$ only some 
 residual, 90\% C.L., allowed regions exist at very small mixing angle, 
$\Delta m^2\approx 10^{-6}-10^{-7}$ eV$^2$ and moderately high 
regular magnetic field [Figure (c)].

The same regions are still acceptable for P=0.7,  
(Figs. \ref{f14N90}). For this level of noise something unexpected 
happens at extremely high 
regular magnetic field [Figure (d)] (probably too high to be acceptable 
on astrophysical grounds): a new, large, 
acceptable region appear for $\Delta m^2\approx 10^{-5}$. 
This region disappears again for extremely 
high chaotic fields (P=0.55, Figure not shown here).
Note that even for this value of P some residual regions with a 
90\% C.L. are marginally acceptable 
from reconciliation of all experiment total rates and 
antineutrino bounds \cite{fio1}
if the regular magnetic field is $\approx 200$ kG
(for $\mu=10^{-11} \mu_B$).

\section{Conclusions}
We have presented calculations of neutrino spin flavor precession in 
presence
a layer of magnetic noise at the bottom of the convective zone 
together with matter and a 
theoretically motivated solar
magnetic field  regular profile. 
We have justified that the level of noise
in this region can   be very high.
We find that MSW regions ($\Delta m^2\approx 10^{-5}$ eV$^2$, both small 
and large mixing solutions) are stable up to very large levels of noise (P=0.7-0.8) but they are acceptable from the point of view of antineutrino production 
only for moderate levels of noise ($P\approx 0.95$).
The stronger r.m.s field occurs at the convective zone, the wider 
$(\delta m^2, \sin^2 2\theta)$ region should be excluded when 
considering  the constrain 
imposed by existing antineutrino bounds.
For strong noise, $P=0.7$ or bigger and reasonable regular magnetic 
field, any parameter region 
$(\Delta m^2, \sin^2 2\theta)$ is excluded. This model of noisy magnetic field 
 is not compatible with particle physics solutions to the SNP. One is allowed then  to 
reverse the problem and to put limits on r.m.s field strength, correlation 
length and transition magnetic moments by demanding a solution to the SNP
 under this scenario.

{\bf Acknowledgments}\\
I  thank my collaborator V.B. Semikoz. 
This work has been supposrted  by DGICYT under Grant 
 PB95-1077 and by  a DGICYT-MEC contract  at Univ. de Valencia.


\begin{thebibliography}{10}


\bibitem{akh4} E. Kh. Akhmedov, Phys. Lett. {B255} (1991) 84. 
\bibitem{fio1}G. Fiorentini, M. Moretti and F.L. Villante, hep-ph/9707097.
\bibitem{fio2}R. Barbieri, G. Fiorentini, G. Mezzorani and  M. Moretti, Phys. Lett. {B259} (1991) 119.
\bibitem{vai1}{ S.I. Vainstein, A.M. Bykov and  I.M. Toptygin,} 
 {\em Turbulence, Current Sheets and Schocks in Cosmic Plasma.} 
Gordon and Breach, 1993.
  \bibitem{vai2}   S.I. Vainshtein, Y.B. Zeldovich and A.A. Ruzmakin, {\em Turbulent
dynamo in Astrophysics}. Nauka, Moscow, 1980.

\bibitem{park1}{ E. N. Parker,}  {\em Cosmical Magnetic Fields.} Clarendon Press, Oxford, 1979.
\bibitem{park2}{ E. N. Parker,}  
Astrophys. J., 408 (1993) 707.

\bibitem{Cataneo}
S.I. Vainstein and F. Cattaneo, 1992, Astrophys. J. 393 (1992) 165.  
\bibitem{Diamond} 
A.V. Gruzinov and P.H. Diamond,  Phys. Rev. Lett.  72 (1994) 1651.
\bibitem{tor3} E. Torrente-Lujan,  Phys. Rev. D59 (1999) 093006. 
in press).
\bibitem{tor4} E. Torrente-Lujan,  Phys. Lett. B441 (1998) 305.


\bibitem{BP95} J.N. Bahcall and M.H. Pinsonneault, Rev. Mod. Phys. {67} (1995) 781.

\bibitem{Ruzmaikin}
A.A. Ruzmaikin, A.M. Shukurov and  D.D. Sokoloff, {\em Magnetic fields of Galaxies}, Kluwer, Dordrecht, 1988.

\bibitem{Zeldovich}
Ya.A. Zeldovich, A.A. Ruzmaikin and D.D. Sokoloff, {\em Magnetic fields in astrophysics}, Gordon and Breach, New York, 1983.


\bibitem{torrente} V.B Semikoz, E. Torrente-Lujan. hep-ph/9809376.
Nucl. Phys. B (In press).


\bibitem{kut1} V.A. Kutvitskii and L.S. Solov'ev, Sov. Phys. JETP. 78 (1994) 456. 

\bibitem{tor2} E. Torrente-Lujan, Phys. Rev. D59 (1999) 073001.

\bibitem{balantekin} F.N. Loreti and A.B. Balantekin, 
Phys. Rev. D50 (1994) 4762.
\bibitem{tor20} E. Torrente-Lujan,  hep-ph/9602398.
\bibitem{bur1} C.P. Burgess and D. Michaud, hep-ph/9606295.
\bibitem{bah2} J.N. Bahcall, P.I. Krastev and  A.Y. Smirnov, hep-ph/9807216.




\end{thebibliography}
\end{document}